\documentclass[
reprint,
floatfix,
amsmath,amssymb,
aps,prapplied,
]{revtex4-2}

\usepackage{amsmath}
\usepackage{amssymb}
\usepackage{graphicx}
\usepackage{dcolumn}
\usepackage{bm}
\usepackage{hyperref}
\usepackage{placeins}
\usepackage{booktabs}
\usepackage{amsthm}
\usepackage{xcolor}

\theoremstyle{plain}

\theoremstyle{definition}

\newcommand{\cc}{{\mathbf{c}}}
\newcommand{\uu}{{\mathbf{u}}}
\newcommand{\xx}{{\mathbf{x}}}
\newcommand{\xm}{{\boldsymbol{x}}}
\newcommand{\zz}{{\mathbf{z}}}
\newcommand{\bb}{{\mathbf{b}}}

\newcommand{\Am}{{\mathbf{A}}}
\newcommand{\CC}{{\mathbf{C}}}
\newcommand{\EE}{{\mathbf{E}}}
\newcommand{\FF}{{\mathbf{F}}}
\newcommand{\GG}{{\mathbf{G}}}
\newcommand{\II}{{\mathbf{I}}}
\newcommand{\JJ}{{\mathbf{J}}}
\newcommand{\HH}{{\mathbf{H}}}

\newcommand{\UU}{{\mathbf{U}}}
\newcommand{\XX}{{\mathbf{X}}}
\newcommand{\Rset}{{\mathbb{R}}}
\newcommand{\Cset}{{\mathbb{C}}}
\newcommand{\dev}[3]{\frac{\text{d}^{#3} #1}{\text{d}#2^{#3}}}

\begin{document}


\title{Physical Reservoir Signal Acquisition for Sub-Nyquist Waveform Reconstruction}

\author{Yuito Ito}
\email{kindaiyuito@stu.kanazawa-u.ac.jp}
\affiliation{%
Graduate School of Natural Science and Technology, Kanazawa University,
Kakuma-machi, Kanazawa, Ishikawa, 920-1192, Japan}

\author{Anas Skalli}
\affiliation{%
Universit\'{e} Marie et Louis Pasteur, CNRS UMR~6174, Institut FEMTO-ST,
15B Avenue Montboucons, Besan\c{c}on, 25000, France}

\author{Tetsuya Asai}
\affiliation{%
Faculty of Information Science and Technology, Hokkaido University,
Kita~14, Nishi~9, Kita-ku, Sapporo, Hokkaido, 060-0814, Japan}

\author{Satoshi Sunada}
\email{sunada@se.kanazawa-u.ac.jp}
\affiliation{%
Faculty of Mechanical Engineering, Institute of Science and Engineering,
Kanazawa University, Kakuma-machi, Kanazawa, Ishikawa, 920-1192, Japan}

\date{\today}

\begin{abstract}
Physical reservoir computing has traditionally exploited the dynamics of physical systems for computational tasks, such as classification and prediction. Here, we introduce a distinct paradigm for exploiting physical reservoirs, termed  \emph{reservoir signal acquisition} (RSA), in which a physical reservoir serves as a dynamical measurement device rather than as a computational engine. In RSA, the reservoir transforms an unknown input waveform into diverse measurements, so that the waveform can be reconstructed from low-rate samples even when its spectral content extends beyond the Nyquist frequency of any individual acquisition channel. The diversity of reservoir dynamics also enables reconstruction of sparse signals with fewer acquisition channels. We experimentally validate RSA using a silicon photonic reservoir device. Through a data-driven calibration requiring no physical model of the device, we reconstruct radio-frequency signals up to 12.5~GHz, four times the Nyquist frequency of each low-rate analog-to-digital converter. Because the required measurement diversity arises from the intrinsic reservoir dynamics, RSA is, in principle, applicable to other physical substrates.
\end{abstract}

\maketitle
\section{Introduction}
\label{sec:intro}

A central theme in modern physics and engineering is the idea that natural physical systems can themselves serve as information processors. Rather than digitally simulating dynamics, one can exploit the intrinsic transformations of a physical substrate, such as optical scattering, mechanical deformation, wave propagation, or nonlinear dynamics, to perform computation directly in the physical layer. This vision has inspired decades of research in unconventional and neuromorphic computing, ranging from physical neural networks~\cite{Wright:2022uv,Momeni:2025vw}, including optical neural networks~\cite{Shastri:2021td}, to brain-inspired hardware architectures~\cite{Mehonic:2022wi,Grollier:2020ur}. The common principle is that computation does not need to be implemented by digital logic: any physical system that performs a useful input--output transformation can become a computing resource.

Reservoir computing~\cite{Jaeger2004,Maass2002,Yan2024ReservoirComputing,Dambre:2012ue,Nakajima2021book} is one of the most general realizations of this principle. A fixed, high-dimensional dynamical system, the so-called \emph{reservoir}, is driven by a time-varying input and generates a rich trajectory of internal states. These states encode diverse temporal representations of the input history, from which a trained linear readout extracts the desired output. Because only the readout is trained, a remarkably broad range of physical systems can serve as reservoirs~\cite{Tanaka:2019uw,Stepney2024,Liang:2024vd}. Experimental demonstrations span delay-line oscillators~\cite{Appeltant:2011vn,Paquot:2012wy,Brunner:2013vh}, memristive devices~\cite{Moon:2019tt,Jang2024Memristive}, soft mechanical systems~\cite{Nakajima:2015wg}, spintronic oscillators~\cite{Torrejon:2017wv}, quantum systems~\cite{Fujii:2017tq}, and optical platforms~\cite{Vandoorne:2014ut,PhysRevX.7.011015,Rafayelyan:2020vh,Sunada:2021wg,Wang:2024wr}. Physical reservoir computing has consequently emerged as a powerful framework for energy-efficient and high-speed temporal processing~\cite{Tanaka:2019uw,PhysRevX.7.011015,Sunada:2020wy,Wang:2024wr,Aadhi:2025vu}.

More recently, reservoir dynamics have gained importance beyond computation. Advances in in-sensor computing \cite{Zhou2020insensor,Wan2023insensor} and neuromorphic sensing \cite{Liu2010neuromorphic,Bartolozzi2022embodied} have motivated the integration of sensing and information processing within a single physical substrate~\cite{Sunada:2019wi,Ito:2026vm,Ehrler2025InSensor,Sun2021insensorRC}, and several studies have explored sensing-oriented reservoir architectures, mainly for inference tasks~\cite{Sun2021insensorRC,Ehrler2025InSensor,Anufriev2024,Anufriev2022}.
Reconstruction of input waveforms or unobserved signals---a fundamental objective in sensing---has also been studied~\cite{Dambre:2012ue,Ito:2026vm,Yamada2026,Zhai2025,Lu2017}; however, whether and under what conditions an unknown waveform can be uniquely recovered when information is lost during the observation process remains unclear for an uncharacterized physical system.

A representative example of such information loss is aliasing in sampling, an indispensable step in measurement because digital processing requires discrete-time data. According to the Nyquist--Shannon theorem~\cite{Shannon:1949uo}, a single analog-to-digital converter (ADC) cannot recover a waveform whose spectral content extends beyond its Nyquist frequency (half the sampling rate). The sampling rate itself cannot be raised arbitrarily without sacrificing resolution~\cite{Walden:1999vi,Murmann2023}; this constraint is distinct from the analog-bandwidth limitation addressed in our previous work~\cite{Ito:2026vm}.

\begin{figure*}[htbp]
\includegraphics[bb=0 0 626 450, width=0.8\linewidth]{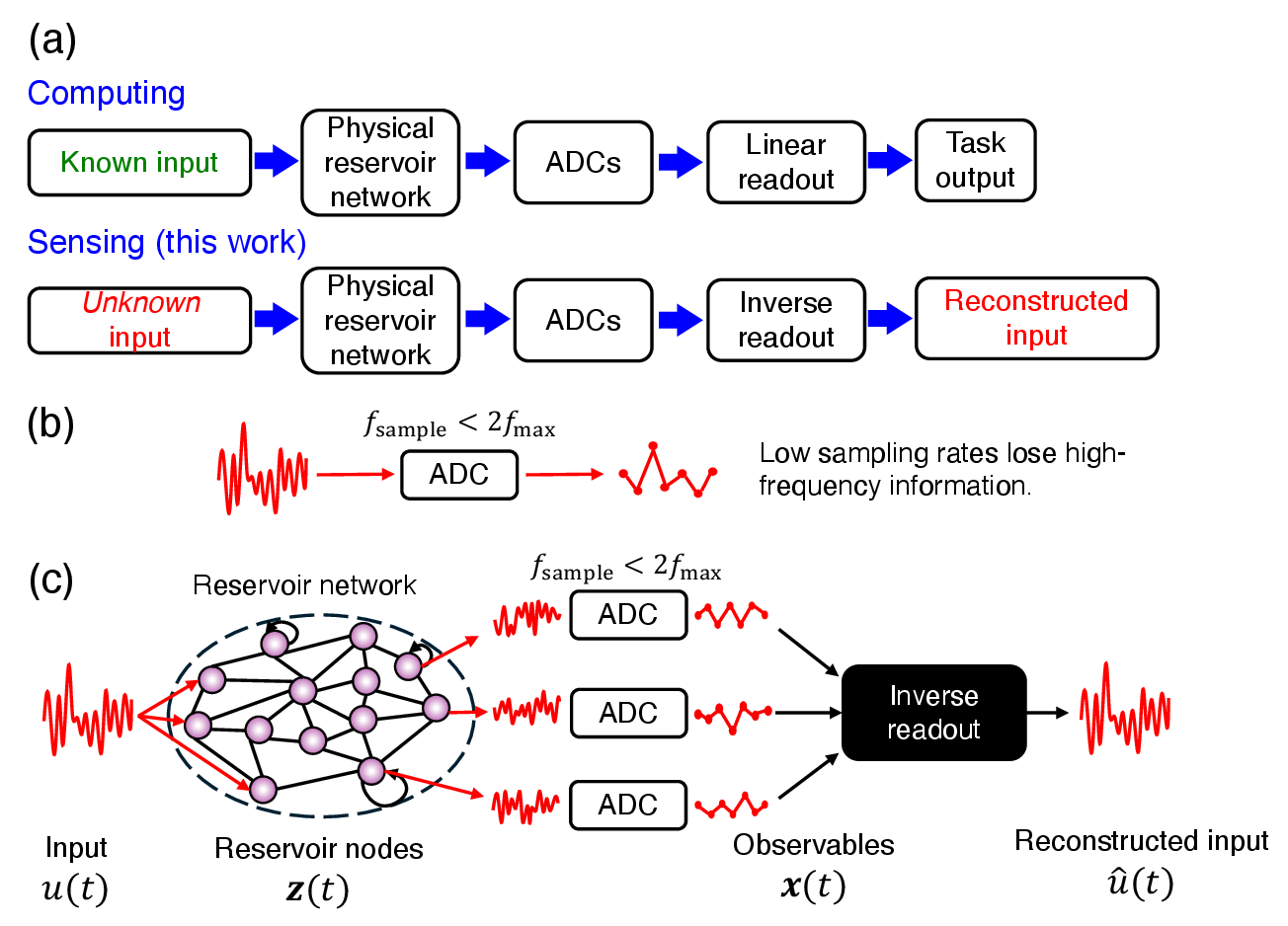}
\caption{Concept of RSA. (a)~Conventional reservoir computing maps a \emph{known} input to a task-specific output via a trained linear readout (top), whereas the present framework reconstructs an \emph{unknown} input from the reservoir response via an inverse readout (bottom). (b)~A single ADC operating below the Nyquist rate ($f_{\rm sample} < 2f_{\rm max}$) cannot recover the input waveform because of aliasing. (c)~RSA overcomes this limitation by combining $M$ distinct reservoir outputs, allowing reconstruction of the input waveform even when every ADC channel samples below the Nyquist rate.}
\label{fig:concept} 
\end{figure*}

In this study, we address this question by using a physical reservoir itself as a dynamical measurement device, a paradigm termed \emph{reservoir signal acquisition} (RSA). Unlike conventional reservoir computing used for inference tasks, RSA reconstructs an unknown input waveform from multiple reservoir outputs with distinct temporal responses, each sampled by a low-rate ADC, thereby overcoming the sampling-rate limit of an individual ADC [Figs.~\ref{fig:concept}(a)--\ref{fig:concept}(c)]. We derive the necessary recovery conditions for general broadband signals and show that sparse signals can be reconstructed using fewer channels.

Importantly, RSA is, in principle, substrate independent: it can be implemented with any physical system exhibiting diverse dynamical responses. In contrast to conventional sub-Nyquist sampling schemes~\cite{Black:1980vh,Razavi:2012vy,Tropp:2010wc,Mishali:2010wy,Eldar2015sampling,Papoulis:1977,Brown:1981}, which rely on a set of designed and precisely characterized measurement channels, in RSA, the required measurement diversity spontaneously emerges from the intrinsic dynamics of a single, uncharacterized physical reservoir. As a result, filter design, prior physical characterization, and a device model are not required. Instead, the inverse mapping is identified entirely from input--output calibration data.

To demonstrate the concept, we experimentally implement RSA using a chaotic-scattering-based silicon photonic reservoir. Using four ADC channels operating at 6.25~GSa/s, we reconstruct various waveforms with frequency components up to 12.5~GHz, corresponding to four times the Nyquist frequency of each ADC.

The remainder of this paper is organized as follows.
Section~\ref{sec:theory} develops the theoretical framework for RSA, including a linear inverse problem formulation and the compressed-sensing-like channel reduction for sparse signals. Section~\ref{sec:simulations} presents the numerical validation. Section~\ref{sec:experiments} describes the silicon photonic experimental platform and waveform reconstruction results. Section~\ref{sec:discussion} provides the discussion. Finally, Section~\ref{sec:conclusion} concludes the paper.

\section{Theoretical Framework}
\label{sec:theory}

\subsection{Reservoir signal acquisition (RSA)}
\label{sec:concept}
Figure~\ref{fig:concept}(c) illustrates the concept of RSA. An unknown input signal drives a physical reservoir, which generates multiple distinct outputs. Each output is sampled by an individual low-rate ADC whose Nyquist frequency is below the maximum frequency of the input; throughout this paper, ``sub-Nyquist'' refers to this per-channel condition. Although no single channel suffices, the input can be reconstructed from all channels together: for general broadband signals, this requires the aggregate sampling rate of all channels to be at least twice the maximum input frequency (Sec.~\ref{sec:inverse}), whereas sparse signals can be recovered at a lower aggregate rate (Sec.~\ref{sec:sparsity}).

To establish the theoretical foundation, we consider a linear sensing model, which captures the essential mechanism of RSA and reduces the problem to a linear inverse problem connected to compressed sensing; the extension to nonlinear reservoirs is presented in Appendix~\ref{sec:nonlinear}.

\subsection{Physical setup and sensing model}
\label{sec:model}

Consider a physical reservoir driven by an input signal $u(t)$, which produces outputs $\{x_m(t)\}_{m=1}^{M}$ through $M$ spatially or modally distinct channels. Let $f_{\rm sample}$ denote the sampling rate of each ADC. In line with the linear framework motivated above, we model each channel by a causal impulse response $h_m(t)$ ($h_m(t)=0$ for $t<0$), such that
\begin{align}
x_m(t) = \int_{-\infty}^{t} h_m(t-t')\,u(t')\,\mathrm{d}t'.
\label{eq:response}
\end{align}
The response functions $\{h_m(t)\}_{m=1}^{M}$ generally differ in their temporal and spectral content owing to the complex internal dynamics of the reservoir. 
The sampling time interval is $\Delta t = 1/f_{\rm sample}$, and each channel is sampled over a time window of duration $T_e$. This yields $N = T_e/\Delta t$ samples per channel during $T_e$. The input signal $u(t)$ has maximum frequency $f_{\rm max}$, and we define the undersampling ratio, 
\begin{align}
R = 2f_{\mathrm{max}}/f_{\mathrm{sample}}.
\end{align}
For $R > 1$, a single ADC channel cannot recover $u(t)$ because of aliasing [Fig.~\ref{fig:concept}(b)]. 

\subsection{Linear inverse problem}
\label{sec:inverse}

We seek to recover $u(t)$ from the $M\times N$ low-rate samples $\{x_m(n\Delta t)\}_{m,n=1}^{M,N}$. To do so, we represent $u(t)$ on a fine time grid with spacing
$\Delta t' = \Delta t/N_R$, where $N_R = \lceil R \rceil$ denotes the ceiling of the undersampling ratio. This grid is sufficiently fine to avoid aliasing; any $f_{\rm max}$-bandlimited signal is uniquely determined by its values on this grid. The total number of unknowns is $N' = N_R N$ (the number of fine-grid samples per window).

Discretizing Eq.~(\ref{eq:response}), the samples at the $m$-th channel satisfy the matrix equation, 
\begin{align}
\xx_m = \HH_m\,\uu,
\label{eq:channel_eq}
\end{align}
where $\uu = \left(u(0), u(\Delta t'), \ldots, u((N'-1)\Delta t')\right)^\top\in \Rset^{N'}$ is the vector of fine-grid input samples, $\xx_m = \left(x_m(\Delta t), x_m(2\Delta t), \ldots, x_m(N\Delta t)\right)^\top \in \Rset^N$ is the vector of measured output samples, and the sensing sub-matrix $\HH_m \in \Rset^{N\times N'}$ has entries ($n=1,\ldots,N$; $n'=0,\ldots,N'-1$)
\begin{align}
\HH_m[n,n'] =
\begin{cases}
h_m(n\Delta t - n'\Delta t')\,\Delta t', &\text{if } n N_R \ge n',\\
0, &\text{otherwise}
\end{cases}
\label{eq:Hm}
\end{align}
The causality condition $n N_R \ge n'$ ensures that only past inputs contribute to each sample.

Because $N < N'$, a single channel is underdetermined: there are more unknowns than equations, which manifests as aliasing. Combining all $M$ channels yields:
\begin{align}
\xx = \HH\,\uu,
\qquad
\HH =
\begin{pmatrix}
\HH_1\\
\HH_2\\
\vdots\\
\HH_M
\end{pmatrix}
\in \Rset^{MN\times N'},
\label{eq:full_system}
\end{align}
where
$\xx = (\xx_1^\top,\ldots,\xx_M^\top)^\top \in \Rset^{MN}$.
The input $\uu$ can then be reconstructed by solving the least-squares problem:
\begin{align}
\hat{\uu}
=
\arg\min_{\uu}
\|\xx-\HH\uu\|^2 .
\end{align}
The solution is uniquely given by the Moore--Penrose pseudoinverse:
\begin{align}
\hat{\uu}
=
\HH^{\dagger}\xx
=
(\HH^\top\HH)^{-1}\HH^\top\xx,
\label{eq:reconstruction}
\end{align}
provided that the combined sensing matrix $\HH$ has full column rank,
\begin{align}
\mathrm{rank}(\HH)=N'.
\end{align}
A necessary condition for full column rank is
\begin{align}
MN \ge N',
\end{align}
or equivalently,
\begin{align}
M \ge N_R.
\label{eq:channel_condition}
\end{align}
In practice, for the random reservoirs considered in Sec.~\ref{sec:simulations}, $\HH$ can have full column rank for $M \ge N_R$ [Fig.~\ref{fig:numexp1}(d)], although its conditioning depends on the diversity of the impulse responses [Fig.~\ref{fig:numexp1}(c)].

We note that the recovery condition is similar to that of the classical generalized sampling theorem of Papoulis and Brown~\cite{Papoulis:1977,Brown:1981}. Whereas classical generalized sampling assumes designed and characterized analysis filters, in RSA, the measurement basis emerges from the intrinsic dynamics of an uncharacterized physical reservoir, and the inverse operator is identified entirely through input--output calibration, as shown in Sec.~\ref{sec:sysid}.

\subsection{Sparsity and channel reduction}
\label{sec:sparsity}
A key distinction from conventional sampling schemes, such as time-interleaved ADCs, emerges when the input signal is sparse. Whereas the channel count of an interleaved ADC is fixed by the desired effective sampling rate, RSA can exploit the measurement diversity to recover sparse signals using substantially fewer channels. 
When $u(t)$ is known to be sparse and representable by only $N'' \ll N'$ basis functions from a dictionary $\Psi \in \Rset^{N'\times N''}$ ($\uu = \Psi\,\mathbf{s}$, $\mathbf{s} \in \Rset^{N''}$), the combined system becomes $\xx = (\HH\Psi)\,\mathbf{s}$, for which $MN \ge N''$ measurements are necessary for unique recovery. The minimum number of channels is reduced to
\begin{align}
M \ge \left\lceil \frac{N''}{N} \right\rceil,
\label{eq:sparse_condition}
\end{align}
which can be substantially smaller than $N_R$ for highly sparse signals $(M < N_R)$. 

\subsection{Data-driven system identification}
\label{sec:sysid}

For a physical reservoir, the impulse responses underlying the sensing matrix $\HH$ are generally unknown and difficult to model accurately. Consequently, obtaining $\HH$ or its pseudoinverse analytically is often impractical. Instead, the reconstruction operator can be identified directly from calibration data.

Using $K$ known pilot signals $\{\uu_k^p\}_{k=1}^{K}$ and their corresponding measurements $\{\xx_k^p\}_{k=1}^{K}$, we form the matrices
\begin{align}
\UU^p=[\uu_1^p\,\cdots\,\uu_K^p],
\qquad
\XX^p=[\xx_1^p\,\cdots\,\xx_K^p].
\end{align}
A least-squares solution of the reconstruction operator is then given by:
\begin{align}
\hat{\HH}^{\dagger}
=
\UU^p{\XX^p}^{\dagger}.
\label{eq:sysid}
\end{align}
This approach requires $K \ge MN$ pilot signals with $\UU^p$ of full row rank $N'$; any rank deficiency of $\XX^p$ (e.g., for $M > N_R$) is handled by the pseudoinverse. 
In practice, using $K \gg MN$ pilots averages out measurement noise and improves the accuracy of $\hat{\HH}^{\dagger}$. Additionally, a row of ones can be appended to $\XX^p$, so that a static offset in the measured signals (Appendix~\ref{sec:photonic_reservoir}) is absorbed into $\hat{\HH}^{\dagger}$.
Because the calibration identifies the end-to-end response of each channel, including its analog front end (photodetector, amplifier, and ADC), systematic effects such as dispersion, optical background, time skew, and channel-to-channel differences in bandwidth and gain are naturally incorporated into $\hat{\HH}^{\dagger}$. Once identified, the same operator can be used to reconstruct any input signal band-limited to $f_{\rm max}$.

\begin{figure*}[htbp]
\includegraphics[bb=0 0 613 498, width=0.7\linewidth]{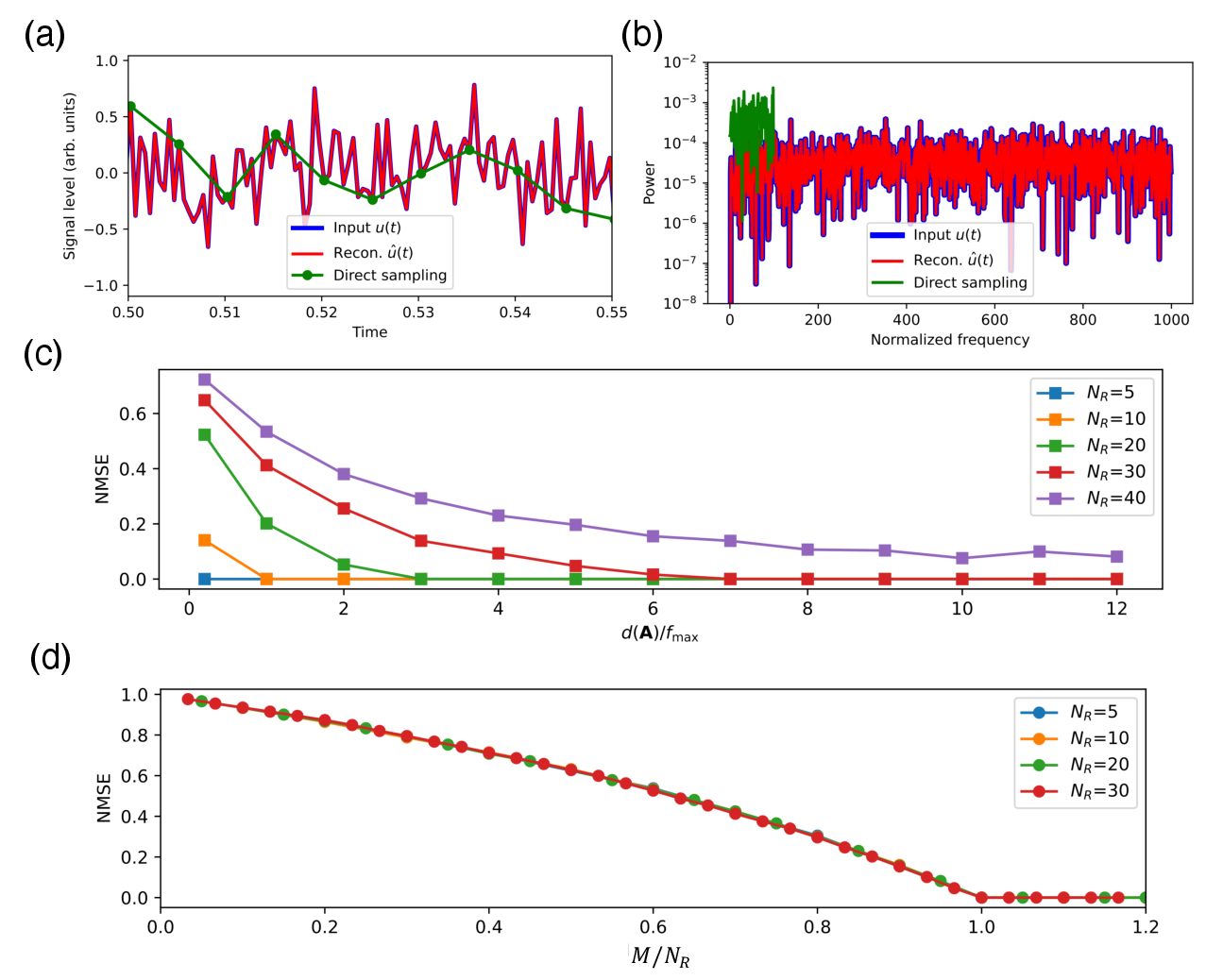}
\caption{Numerical validation of RSA for a broadband random waveform $u(t)$ with $f_{\rm max} = N_R\,(f_{\rm sample}/2)$. (a)~Time-domain comparison of $u(t)$ (blue), the waveform directly sampled by a single ADC (green), and the reconstructed waveform $\hat{u}(t)$ (red) obtained by the RSA method with $M = N_R = 10$ ADCs. (b)~Corresponding power spectra on a logarithmic amplitude scale. (c)~Reconstruction NMSE as a function of the normalized eigenvalue spread $d(\Am)/f_{\rm max}$ for $M=N_R=5$, $10$, $20$, $30$, and $40$. (d)~Reconstruction NMSE as a function of the normalized channel count $M/N_R$ for $N_R=5$, $10$, $20$, and $30$, with $d(\Am)/f_{\rm max}=10$.}
\label{fig:numexp1}
\end{figure*}

\section{Numerical Simulations}
\label{sec:simulations}

To validate the theoretical framework developed in Sec.~\ref{sec:theory}, we consider an $M'$-dimensional linear reservoir network with $M$ output channels [Fig.~\ref{fig:concept}(c)], described by the state-space model
\begin{align}
\dev{\zz(t)}{t}{} &= \Am\,\zz(t) + \bb\, u(t), \label{eq:state}\\
\xm(t)     &= \CC\,\zz(t), \label{eq:output}
\end{align}
where $\zz(t) \in \Rset^{M'}$ denotes the internal state of the reservoir, $\Am \in \Rset^{M'\times M'}$ is the system matrix, $\bb \in \Rset^{M'}$ is the input coupling vector, and $\CC \in \Rset^{M\times M'}$ is the output matrix that maps the state to the $M$ output channels, each sampled at rate $f_{\rm sample}$.

In this simulation, the entries of $\Am$ and $\bb$ are drawn i.i.d.\ from a uniform distribution, with all the eigenvalues of $\Am$ constrained to have negative real parts to ensure stability. The eigenvalue spread $d(\Am) = \max_{m'}|\lambda_{m'}| - \min_{m'}|\lambda_{m'}|$, where $\lambda_{m'} \in \Cset$ ($m' = 1, \ldots, M'$) are the eigenvalues of $\Am$, characterizes the range of the reservoir's characteristic frequencies. We set $M' = 500 \gg M$.

For a broadband random input $u(t)$ with $f_{\mathrm{max}} = N_R\,(f_{\mathrm{sample}}/2)$ and $N_R = 10$, the waveform sampled by a single ADC is severely distorted by aliasing [Fig.~\ref{fig:numexp1}(a)], and its spectrum folds back into the Nyquist band [Fig.~\ref{fig:numexp1}(b)]. In contrast, the pseudoinverse reconstruction [Eq.~(\ref{eq:reconstruction})] from $M = N_R = 10$ reservoir outputs recovers the input with high fidelity in both the time and frequency domains in this noise-free simulation.

Figure~\ref{fig:numexp1}(c) shows the normalized mean squared error (NMSE) as a function of the normalized eigenvalue spread $d(\Am)/f_{\mathrm{max}}$. The NMSE decreases overall with $d(\Am)/f_{\mathrm{max}}$, and for $N_R \le 30$, accurate reconstruction is achieved once the eigenvalue spread is sufficiently larger than $f_{\rm max}$, because a wide spread of reservoir modes increases the observability of the high-frequency signal components. A larger $N_R$ requires a larger eigenvalue spread to achieve the same accuracy.

Figure~\ref{fig:numexp1}(d) shows the NMSE as a function of the normalized channel count $M/N_R$. The NMSE decreases with $M/N_R$ and vanishes at $M/N_R \approx 1$; when $M < N_R$, the sensing matrix $\HH$ does not have full column rank, whereas once $M \ge N_R$, the random reservoirs considered here yield a full-column-rank sensing matrix. The curves for all values of $N_R$ collapse onto a single curve, reflecting the channel condition $M \ge N_R$ in Eq.~(\ref{eq:channel_condition}).

\begin{figure*}[htbp]
\includegraphics[bb=0 0 720 376, width=0.6\linewidth]{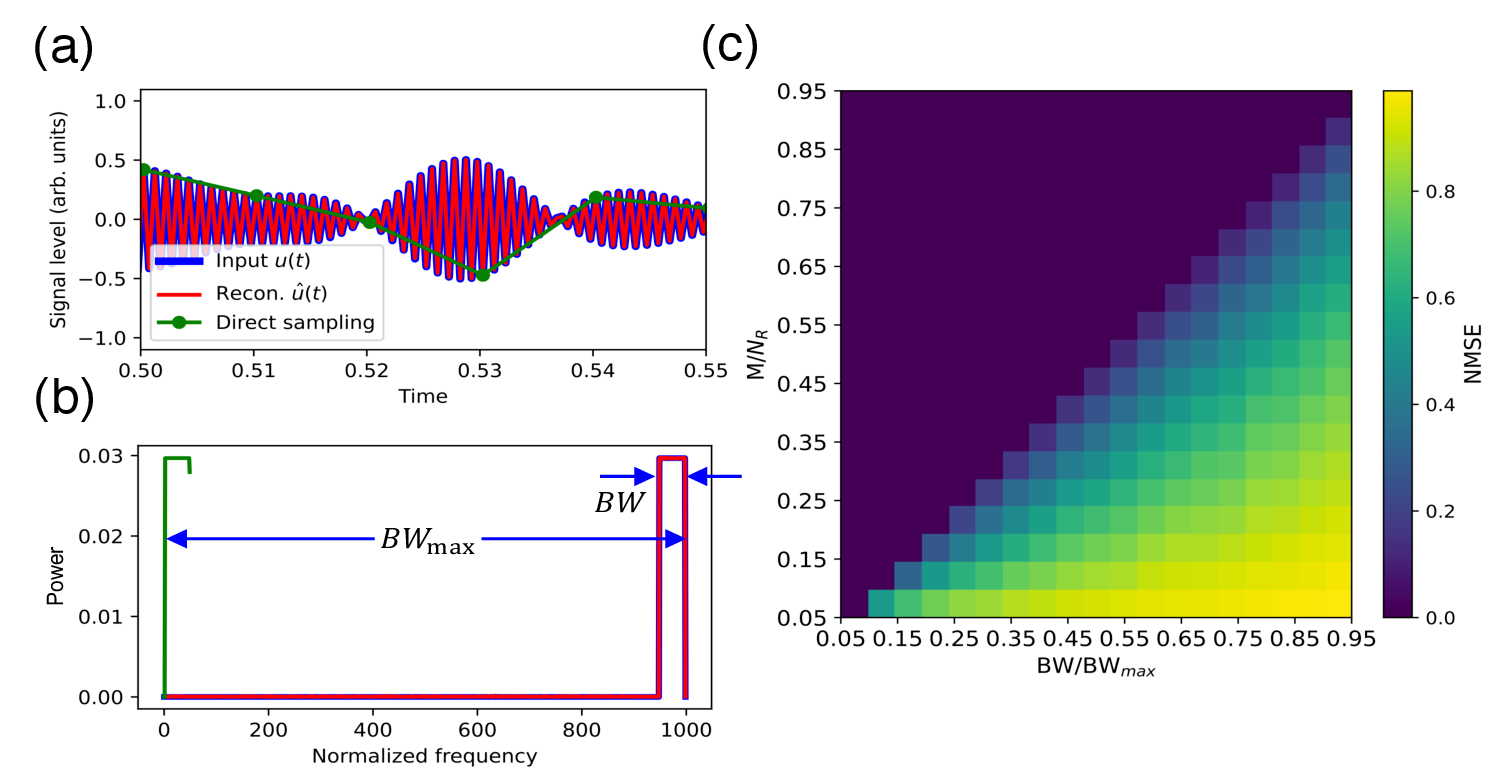}
\caption{\label{fig:numexp2}
Channel reduction for spectrally confined signals with fractional bandwidth $BW/BW_{\rm max}$. (a)~Time-domain comparison of the input waveform (blue), the single-ADC output (green), and the reconstruction (red) obtained when $M = 1$ and $N_R = 10$ for $d(\Am)/f_{\rm max} = 10$. (b)~Corresponding power spectra showing the signal bandwidth $BW$ relative to the maximum measurable bandwidth $BW_{\rm max}$. (c)~NMSE as a function of $M/N_R$ and  $BW/BW_{\rm max}$; the boundary of the dark region approximately follows $M/N_R \approx BW/BW_{\rm max}$.
}
\end{figure*}

Figure~\ref{fig:numexp2} verifies the channel reduction predicted by Eq.~(\ref{eq:sparse_condition}) for spectrally confined inputs with a known bandwidth $BW$ narrower than the maximum measurable bandwidth $BW_{\rm max} = f_{\mathrm{max}}$; the pilot signals are confined to the same bandwidth. Figures~\ref{fig:numexp2}(a) and \ref{fig:numexp2}(b) show a representative reconstruction using only a single channel ($M = 1 < N_R = 10$) in the time and frequency domains, respectively. Figure~\ref{fig:numexp2}(c) presents the NMSE as a function of $M/N_R$ and the fractional bandwidth $BW/BW_{\rm max}$, corresponding to $N''/N'$. Accurate reconstruction (dark region) is achieved for $M/N_R \gtrsim BW/BW_{\rm max}$, in agreement with Eq.~(\ref{eq:sparse_condition}).

\section{Experiments}
\label{sec:experiments}
\subsection{Experimental setup}
\label{sec:setup}

\begin{figure*}[htbp]
\includegraphics[bb=0 0 700 246, width=0.75\linewidth]{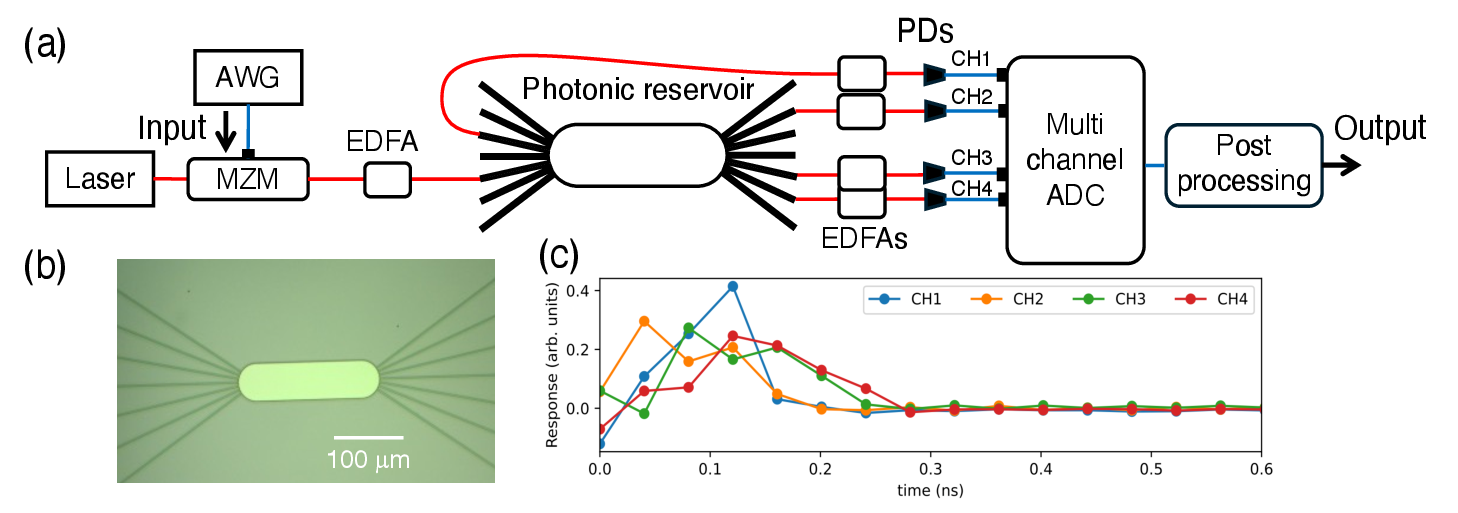}
\caption{(a)~Schematic of the experimental setup. AWG: arbitrary waveform generator; MZM: Mach--Zehnder modulator; EDFA: erbium-doped fiber amplifier; PDs: photodetectors; ADC: analog-to-digital converter. (b)~Micrograph of the silicon photonic reservoir based on chaotic scattering in a stadium-shaped cavity. Adapted with permission from Ref.~\cite{Ito:2026vm} \copyright~Optica Publishing Group. (c)~Measured impulse responses $h_m(t)$ at four selected output channels (CH1--CH4).}
\label{fig_setup}
\end{figure*}

The experimental setup is shown in Fig.~\ref{fig_setup}(a). The input waveform $u(t)$, generated by an arbitrary waveform generator (Tektronix AWG70002B, 25~GSa/s), is imprinted onto the output of a continuous-wave laser (Alnair Labs TLG-220, linewidth $<100$~kHz) by a Mach--Zehnder intensity modulator (Exail MXAN-LN-10, 3-dB bandwidth $\sim12$~GHz). After amplification using an erbium-doped fiber amplifier (Thorlabs EDFA100P), the modulated light is injected into the silicon photonic reservoir. The outputs from $M$ selected reservoir ports are optically amplified and detected by wideband InGaAs photodetectors (Thorlabs RXM25BF, 500~kHz--25~GHz). The detected signals are digitized by a real-time 8-bit oscilloscope (Tektronix DPO72504DX, analog bandwidth 23~GHz in four-channel mode) at $f_{\rm sample}=6.25$~GSa/s per channel, i.e., with a Nyquist frequency of 3.125~GHz. For a maximum signal frequency of $f_{\rm max}=12.5$~GHz, the undersampling ratio is $N_R=4$. The pseudoinverse $\hat{\HH}^{\dagger}$ is calibrated from $K\approx 7000$ broadband random pilot signals ($N'=200$ samples) using Eq.~(\ref{eq:sysid}). 

\paragraph{Silicon photonic reservoir.}

The reservoir used in the experiment is a stadium-shaped optical microcavity fabricated on a silicon photonic chip [Fig.~\ref{fig_setup}(b)]. The same type of reservoir was used in Ref.~\cite{Ito:2026vm} to extend the analog detection bandwidth of photodetectors; here, it is used to overcome the sampling-rate limit of the ADCs. The cavity is coupled to 14 single-mode waveguides, comprising one input port and 13 output ports located at distinct positions along the cavity boundary. The classically chaotic ray dynamics of the stadium geometry ensure that each output channel probes a different superposition of the cavity modes \cite{RevModPhys8761,Harayama:2011uf}, yielding diverse impulse responses $\{h_m(t)\}_{m=1}^M$ with a decay time of a few hundred picoseconds [Fig.~\ref{fig_setup}(c)].

We note that the reservoir outputs are measured as optical intensities, resulting in a nonlinear input--output relationship. However, for a small modulation depth and the short impulse responses shown in Fig.~\ref{fig_setup}(c), the measured intensity is approximately linear in $u(t)$, so that the linear inverse model of Sec.~\ref{sec:theory} applies (Appendix~\ref{sec:photonic_reservoir}).

\subsection{Broadband waveform reconstructions}
\label{sec:results}

\begin{figure*}[htbp]
\includegraphics[bb=0 0 666 505, width=0.75\linewidth]{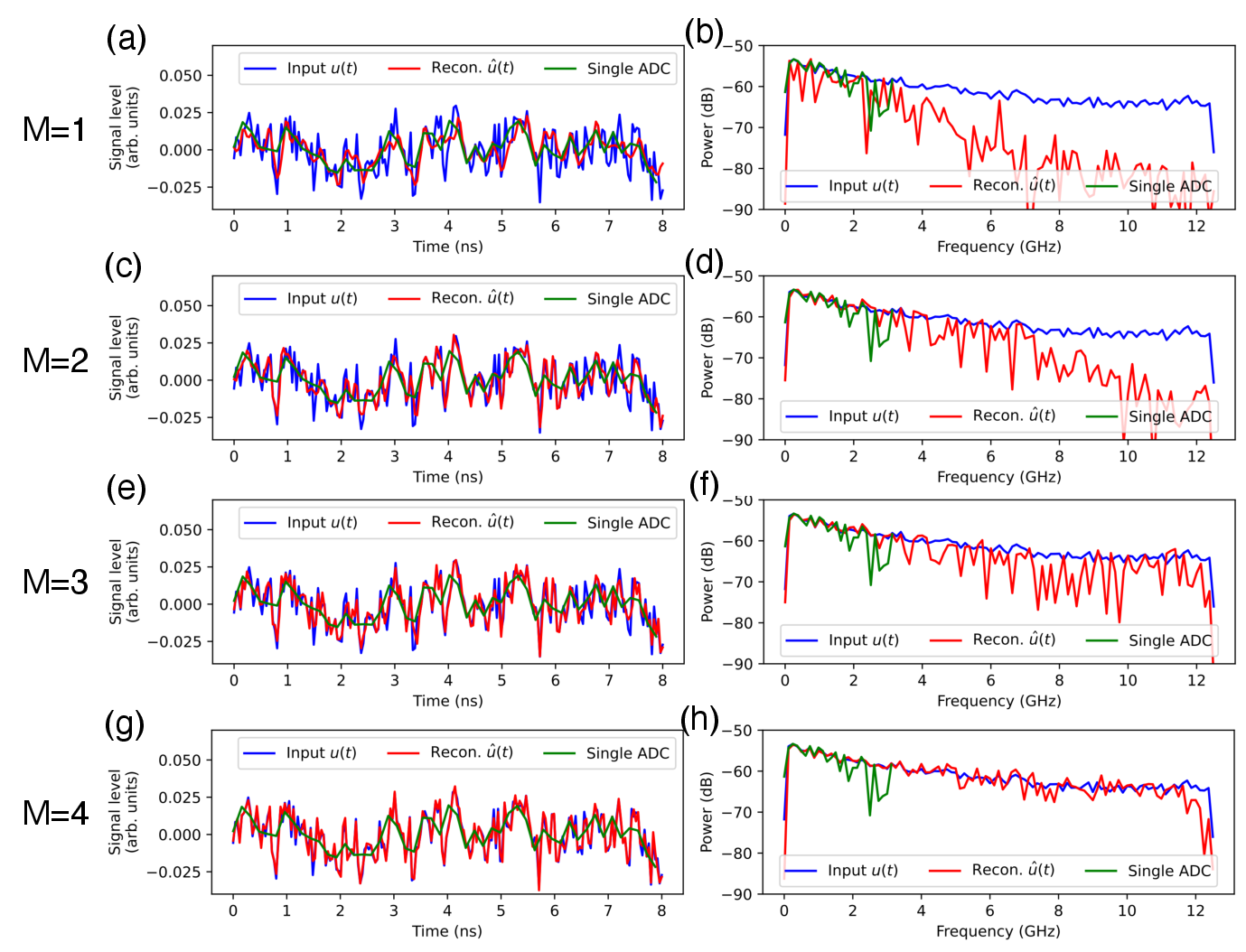}
\caption{Reconstruction of a 500 kHz--12.5~GHz broadband random waveform ($N_R=4$) using $M=1$--$4$ ADC channels. Left: input waveform $u(t)$ (blue), reconstruction $\hat{u}(t)$ (red), and single-ADC output (green). Right: corresponding power spectra.}
\label{fig_rand}
\end{figure*}

\begin{figure}[htbp]
\includegraphics[bb=0 0 245 173, width=0.6\linewidth]{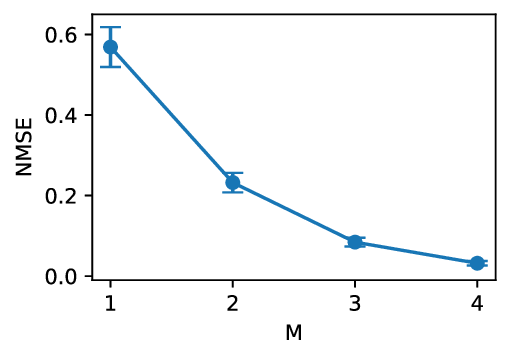}
\caption{NMSE as a function of $M$. Symbols and error bars denote the mean and standard deviation over 30 broadband random waveforms.}
\label{fig_nmse_m}
\end{figure}

Figure~\ref{fig_rand} shows the reconstruction of a 500 kHz--12.5~GHz broadband random waveform, not used for calibration, using $M=1$--$4$ ADC channels, together with their corresponding spectra. For $M=1$, the reconstructed waveform captures only limited features of the original signal. As $M$ increases, both the time-domain waveform and spectral estimate become progressively more accurate. 
Figure~\ref{fig_nmse_m} shows the NMSE evaluated over 30 broadband random waveforms. The NMSE decreases monotonically from $0.57\pm0.05$ at $M=1$ to $0.032\pm0.006$ at $M=N_R=4$. As discussed in Appendix~\ref{sec:ampnoise}, redundant channels ($M > N_R$) improve the conditioning of the sensing matrix and further decrease the NMSE~\cite{Supplement}.

Because the response of the chaotic cavity depends sensitively on the wavelength, operating the chip at different wavelengths provides different reservoir realizations, for which similar NMSEs were obtained with a separate calibration for each wavelength~\cite{Supplement}.

\subsection{Reconstruction of various waveforms}
\label{sec:waves}
\begin{figure*}[htbp]
\includegraphics[bb=0 0 644 289, width=0.75\linewidth]{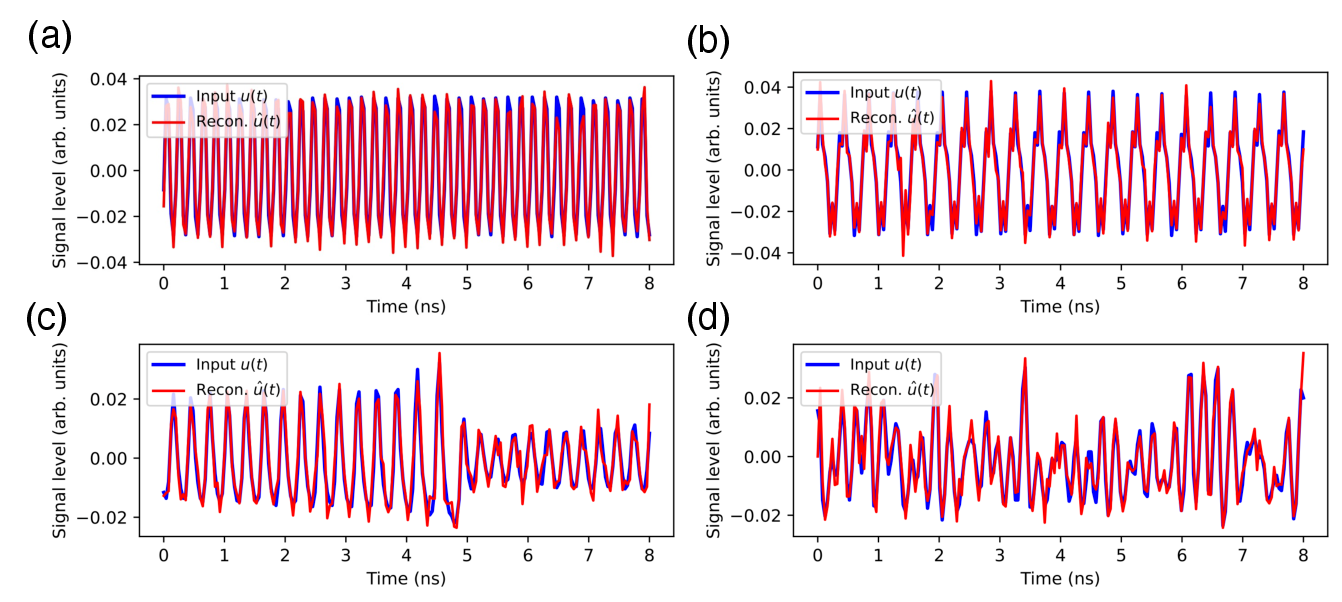}
\caption{Reconstruction of representative waveform classes using $M=4$. (a) 5.0-GHz sinusoid. (b) Two-tone signal with identical alias frequency. (c) Santa~Fe chaos. (d) Lang--Kobayashi laser chaos. Blue: input waveform. Red: reconstruction.}
\label{fig_waves}
\end{figure*}

The operator calibrated with broadband random pilot signals (Sec.~\ref{sec:results}) can be applied directly to reconstruct other classes of waveforms.
Figure~\ref{fig_waves} presents reconstruction results for four qualitatively different waveform classes using $M=4$: (a) a 5.0-GHz sinusoid, (b) a two-tone signal whose components alias to the same baseband frequency in a single ADC, (c) a chaotic waveform from the Santa~Fe competition dataset~\cite{Weigend1994}, and (d) a broadband Lang--Kobayashi laser-chaos waveform~\cite{uchida2012optical}. The corresponding NMSEs are (a) 0.028, (b) 0.025, (c) 0.084, and (d) 0.083.

\subsection{PAM4 signal reconstruction}
\label{sec:pam4}
\begin{figure}[htbp]
\includegraphics[bb=0 0 648 477, width=\linewidth]{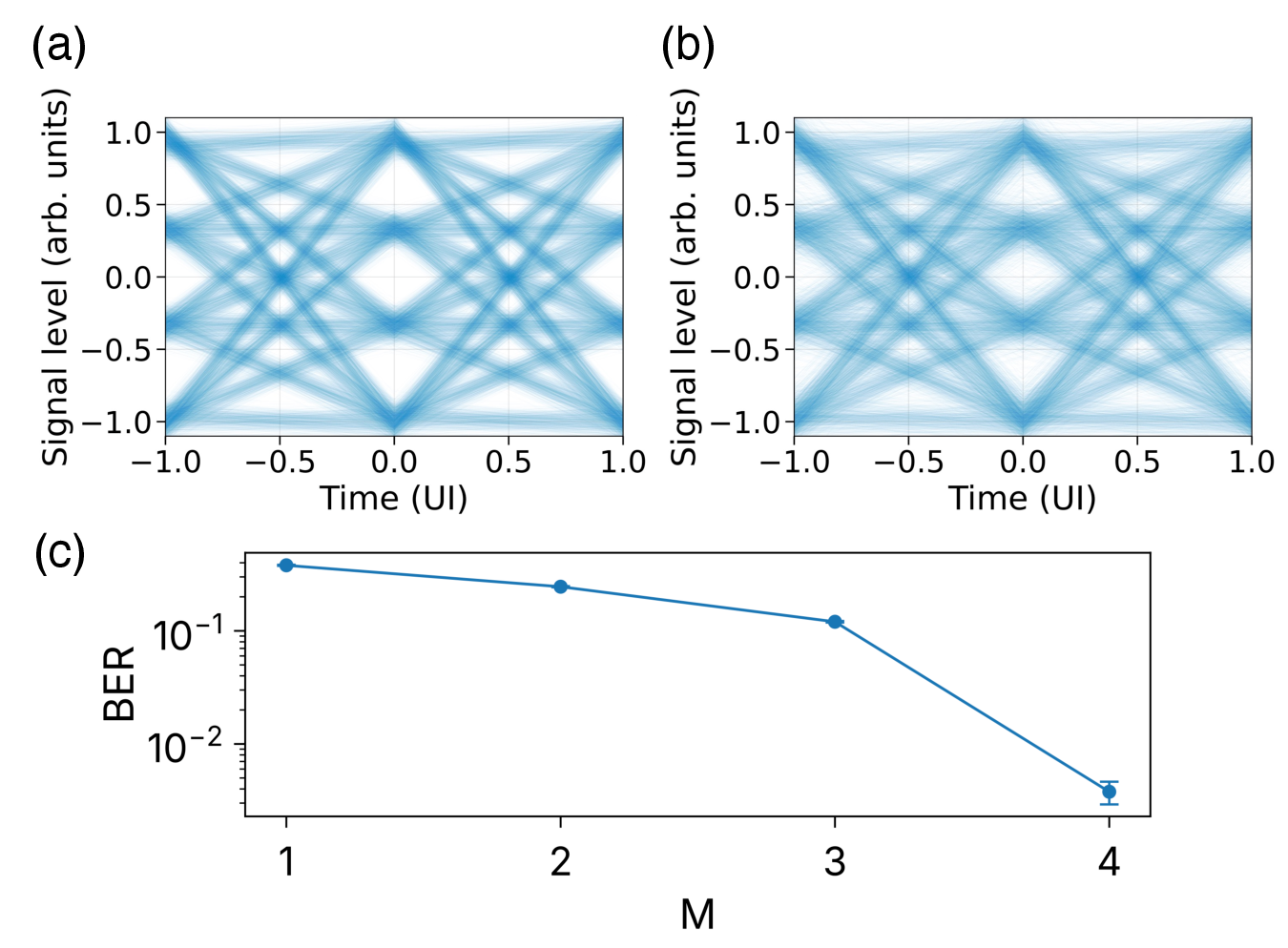}
\caption{Reconstruction of a 25-GBd PAM4 signal. Eye diagrams of (a) the MZM output sampled at 25~GSa/s with FIR equalization and (b) the RSA reconstruction ($M=4$, 6.25~GSa/s per channel). In (a,b), adjacent samples are connected by straight lines. UI: unit interval. (c) BER versus $M$ over all $1.8\times10^{4}$ bits. Error bars denote the standard error of the mean over 45 segments of 200 symbols~\cite{Supplement}.}
\label{fig:pam4}
\end{figure}

We also reconstructed a 25-GBd four-level pulse-amplitude-modulation (PAM4) signal (9000 symbols), whose symbol rate is four times the sampling rate of each ADC. The operator was calibrated with broadband random pilot signals in the same manner. As a reference, the MZM output was directly sampled at 100~GSa/s, downsampled to 25~GSa/s, and equalized with a 15-tap finite-impulse-response (FIR) filter trained on the same pilot signals [Fig.~\ref{fig:pam4}(a)]. The RSA reconstruction with $M=4$ shows four clearly distinguishable levels [Fig.~\ref{fig:pam4}(b)], with a bit error rate (BER) of $3.8\times10^{-3}$, below a typical soft-decision forward-error-correction threshold of approximately $2\times10^{-2}$, whereas no errors were detected for the reference among $1.8\times10^{4}$ bits. As shown in Fig.~\ref{fig:pam4}(c), the BER decreases by more than an order of magnitude from $M=3$ to $M=N_R=4$, consistent with Eq.~(\ref{eq:channel_condition}).

\subsection{Sparse signal reconstruction}
\begin{figure}[htbp]
\includegraphics[bb=0 0 298 361, width=0.75\linewidth]{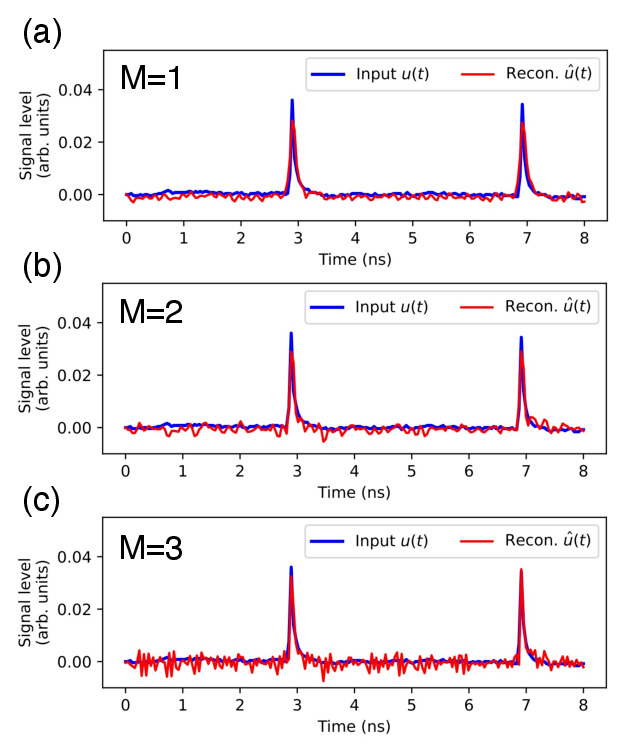}
\caption{Reconstruction of 100-ps short pulses using (a) $M=1$, (b) $M=2$, and (c) $M=3$ ($N_R=4$). NMSEs are (a) 0.263, (b) 0.218, and (c) 0.293, respectively.}
\label{fig_sparse} 
\end{figure}

Figure~\ref{fig_sparse} shows reconstruction results for a short pulse train with a 100-ps pulse width, which is sparse in the temporal domain. Even at $M = 1 < N_R = 4$ and without prior knowledge of the pulse positions, the pulses are reconstructed with an NMSE of 0.263, suggesting that the short, localized impulse responses $\{h_m(t)\}$ make $\hat{\HH}^{\dagger}$ act effectively as a local deconvolution operator, which favors temporally localized (sparse) signals even without an explicit sparsity constraint. However, the NMSE remains at 0.2--0.3 for all $M$. Because the 100-ps pulses occupy only a small fraction of the window, the signal energy that normalizes the NMSE is small, so that the noise floor between the pulses dominates the NMSE even though the pulses themselves are reproduced for all $M$.

\section{Discussion}
\label{sec:discussion}

The RSA demonstrated in this study expands the role of physical reservoirs from dynamical processors to dynamical measurement devices, complementing rather than replacing the conventional reservoir-computing paradigm. The proposed framework has three distinctive features. 
First, the measurement basis required for sub-Nyquist reconstruction arises naturally from the high-dimensional fading-memory dynamics of the physical substrate itself; unlike engineered filter banks, it requires no filter design or parameter optimization.
In contrast to speckle-based photonic compressive sensing~\cite{Borlaug:2021}, in which a wavelength-dependent speckle pattern is imprinted on chirped optical pulses and signal sparsity is required, RSA exploits the temporal responses of the reservoir and can also reconstruct non-sparse broadband waveforms.

Second, the measurement diversity of the reservoir reduces the number of required acquisition channels for sparse signals.

Third, the distributed multi-channel encoding can provide inherent robustness to timing imperfections and amplitude noise. The channels require a common frequency reference, but no phase alignment among their sampling clocks (Appendix~\ref{sec:clockskew}). Because each channel observes a physically distinct filtered version of the input with uncorrelated jitter, the upper bound of the mean squared reconstruction error due to random jitter scales as $1/M$ for $M \gg N_R$ (Appendix~\ref{sec:jitter}). Likewise, the calibrated operator weights each channel by its SNR, so noisier channels are down-weighted (Appendix~\ref{sec:ampnoise}); the remaining SNR cost relative to a Nyquist-rate ADC is determined by the conditioning of the sensing matrix.

The reconstruction up to four times the Nyquist frequency of each ADC demonstrated here is not an intrinsic limit of the approach: as indicated in Fig.~\ref{fig:numexp1}(c), accurate reconstruction requires a large spectral spread of the impulse responses. A smaller or more dispersive cavity is therefore a direct route toward a larger spectral spread. 
On the acquisition side, the analog front end of each channel must still pass frequencies up to $f_{\rm max}$; RSA relaxes the sampling rate of each channel, not its analog bandwidth. Such front ends are available, for example, in RF-sampling ADCs, whose analog input bandwidth considerably exceeds their Nyquist frequency~\cite{TIADC}. These design strategies may enable waveform acquisition beyond the sampling-rate limits of current electronic digitizers, for example, in high-speed optical communications and coherent optical metrology.

Although the theoretical framework in Secs.~\ref{sec:theory} and~\ref{sec:simulations} is formulated for linear reservoirs, a nonlinear reservoir can, in principle, also serve as the measurement device, provided that an appropriate inverse mapping is identified from calibration data (Appendix~\ref{sec:nonlinear}).

For implementation outside the laboratory, the dependence of the reservoir response on temperature and wavelength must be taken into account. The reconstruction operator can be maintained by periodic recalibration with pilot signals or by adaptive tracking, analogous to the adaptive equalizers used in coherent optical receivers.

\section{Conclusion}
\label{sec:conclusion}

We demonstrated reservoir signal acquisition (RSA), a paradigm in which a physical reservoir acts as a dynamical measurement device for sub-Nyquist waveform acquisition. Using a silicon photonic chaotic-cavity reservoir with four ADC channels at 6.25~GSa/s, we reconstructed broadband waveforms up to 12.5~GHz, four times the Nyquist frequency of each ADC, as well as a 25-GBd PAM4 signal. In principle, RSA can be implemented with other physical substrates, exploiting the dynamical richness of physical reservoirs as an alternative to high-sampling-rate digitizers.

\begin{acknowledgments}
This work was supported by JSPS KAKENHI (Grant Nos.\ JP22H05198,
JP23K28157, JP26K21733), and Japan Science and Technology Agency (JST),
Core Research for Evolutional Science and Technology (CREST)
(Grant No.\ JPMJCR24R2). During the preparation of this manuscript, the authors used Claude Opus 5.5 (Anthropic) to proofread the manuscript and improve its clarity and readability. The suggestions generated by the AI tool were reviewed and edited as necessary by the authors, who take full responsibility for the final content of this article.
\end{acknowledgments}

\section*{DATA AVAILABILITY}
The data that support the findings of this article are available from
the corresponding author upon reasonable request.


\appendix

\section{Waveform reconstruction with nonlinear reservoirs}
\label{sec:nonlinear}

For nonlinear reservoir systems with fading memory, the nonlinear input--output relationship can be systematically expanded via a Volterra series for sufficiently small inputs \cite{Boyd1985,rugh1981nonlinear}. Truncating this expansion, the reconstruction of input waveforms can be formulated as a structured nonlinear inverse problem.

\subsection{Volterra series representation}

We consider a reservoir governed by a nonlinear state equation and an observation function,
\begin{align}
\frac{\mathrm{d}\zz}{\mathrm{d}t} = \FF(\zz(t),u(t)),
\qquad
\xm(t) = \GG(\zz(t)),
\label{eq:nl_state}
\end{align}
where $\zz(t)\in\mathbb{R}^{M'}$ is the reservoir state, $\FF:\mathbb{R}^{M'}\times\mathbb{R}\to\mathbb{R}^{M'}$ is the vector field, and $\GG:\mathbb{R}^{M'}\to\mathbb{R}^{M}$ is the observation map.

Assume that the reservoir defines a causal fading-memory input--output operator and that $\FF$ and $\GG$ are sufficiently smooth (e.g., locally analytic) around the unforced equilibrium such that, for sufficiently small inputs $u$, the output admits a locally convergent Volterra series \cite{Boyd1985,rugh1981nonlinear}:
\begin{align}
&x_m(t) =  \nonumber \\
&\sum_{n=0}^{\infty}
\int_0^\infty\!\!\cdots\!\!\int_0^\infty
h_m^{(n)}(\tau^{(1)},\ldots,\tau^{(n)})
\prod_{l=1}^n u(t-\tau^{(l)})\,
\mathrm{d}\tau^{(1)}\cdots\mathrm{d}\tau^{(n)},
\label{eq:volterra}
\end{align}
where $h_m^{(n)}$ is the $n$th-order Volterra kernel of the $m$th output channel, and the $n=0$ term represents a constant offset. The $n=1$ term reduces to the linear convolution model in Eq.~(\ref{eq:response}), whereas the higher-order terms capture nonlinear responses.

\subsection{Second-order truncation and inverse problem formulation}

Truncating Eq.~(\ref{eq:volterra}) at second order, which captures the leading nonlinear correction, gives
\begin{align}
&x_m(t)
\approx
h_m^{(0)}
+ 
\int_0^\infty h_m^{(1)}(\tau^{(1)})\,u(t-\tau^{(1)})\,\mathrm d\tau^{(1)} \nonumber \\
&+
\int_0^\infty \int_0^\infty
h_m^{(2)}(\tau^{(1)},\tau^{(2)})\,
u(t-\tau^{(1)})u(t-\tau^{(2)})\,
\mathrm d\tau^{(1)}\mathrm d\tau^{(2)}.
\end{align}
Discretizing the input on the fine grid $u_r = u(r\Delta t')$ $(r=0,\dots,N'-1)$ and sampling the output at times $t_k$ $(k=1,\dots,N)$, we obtain
\begin{align}
x_m(t_k)
&\approx
c_{m}
+
\sum_{r=0}^{N'-1} H^{(1)}_{m,kr}u_r \nonumber \\
&+
\sum_{r=0}^{N'-1}\sum_{s=0}^{N'-1}
H^{(2)}_{m,krs}\,u_r u_s,
\end{align}
where the coefficients $H^{(1)}_{m,kr}$ and $H^{(2)}_{m,krs}$ absorb the discretized Volterra kernels along with the quadrature weights. Collecting the outputs into
$\xx_m=(x_m(t_1),\dots,x_m(t_N))^\top\in\mathbb{R}^N$
and the input samples into
$\uu=(u_0,\dots,u_{N'-1})^\top \in \mathbb{R}^{N'},$
and matricizing the index pair $(r,s)$ of $H^{(2)}_{m,krs}$, the above expression can be written compactly as
\begin{align}
\xx_m
\approx
\cc_m+\HH_m^{(1)}\uu+\HH_m^{(2)}(\uu\otimes\uu),
\label{eq:volterra_disc} 
\end{align}
where 
$\cc_m \in \Rset^{N}$ is the zeroth-order (DC) vector, $\HH_m^{(1)} \in \Rset^{N\times N'}$ and $\HH_m^{(2)} \in \Rset^{N\times N'^2}$ are the discretized first- and second-order Volterra kernel matrices, and $\uu\otimes\uu=\mathrm{vec}(\uu\uu^\top)\in\mathbb{R}^{N'^2}$. Defining the degree-2 feature vector 
\begin{align} 
\bm{\varphi}(\uu) = \bigl(1,\;\uu^\top,\;(\uu\otimes\uu)^\top\bigr)^\top \in \Rset^{P}, 
\quad 
P = 1+N'+N'^2, 
\label{eq:feature_vec} 
\end{align} 
Eq.~(\ref{eq:volterra_disc}) is written compactly as $\xx_m = \JJ_m\,\bm{\varphi}(\uu)$, where $\JJ_m = [\cc_m,\,\HH_m^{(1)},\,\HH_m^{(2)}] \in \Rset^{N\times P}$. Stacking all $M$ channels: 
\begin{align} 
\xx = \JJ\,\bm{\varphi}(\uu), \qquad 
\JJ = 
\begin{pmatrix}
\JJ_1\\ \vdots \\ \JJ_M
\end{pmatrix} 
\in \Rset^{MN\times P}. 
\label{eq:nl_system} 
\end{align} 
Given $\JJ$, the waveform reconstruction reduces to the nonlinear inverse problem 
\begin{align} 
\hat{\uu} = \arg\min_{\uu \in \Rset^{N'}}\,\bigl\|\xx - \JJ\,\bm{\varphi}(\uu)\bigr\|^2. \label{eq:nl_inverse} 
\end{align} 
Because $\bm{\varphi}(\uu)$ is nonlinear in $\uu$, Eq.~(\ref{eq:nl_inverse}) is generally nonconvex; iterative solvers such as gradient descent or the Levenberg--Marquardt method can be applied to obtain $\hat{\uu}$. The same construction extends directly to any finite truncation order $p\ge 3$ by augmenting the feature vector with higher-order Kronecker powers $\uu^{\otimes n}$ $(n=3,\dots,p)$ and incorporating the corresponding matricized Volterra kernels into $\JJ$. In this case, the reconstruction problem retains the same compact form $\xx=\JJ\bm{\varphi}_p(\uu)$ and is reduced to a nonlinear least-squares problem of the form in Eq.~(\ref{eq:nl_inverse}).

\subsection{Data-driven system identification}
\label{sec:nl_sysid}

The kernel matrix $\JJ$ is generally unknown and is estimated
from $K$ pilot pairs $\{(\uu_k^p,\,\xx_k^p)\}_{k=1}^K$.
Forming the pilot matrices
$\XX^p = [\xx_1^p,\ldots,\xx_K^p]$ and
$\bm{\Phi}^p = [\bm{\varphi}(\uu_1^p),\ldots,\bm{\varphi}(\uu_K^p)]$,
the least-squares estimate is
\begin{align}
\hat{\JJ} = \XX^p\,{\bm{\Phi}^{p}}^{\!\dagger},
\label{eq:nl_sysid}
\end{align}
in direct analogy with the linear calibration of Eq.~(\ref{eq:sysid}).
Because $\uu\otimes\uu$ is symmetric, only $N'(N'+1)/2$ of its entries are independent. Replacing $\uu\otimes\uu$ by the half-vectorization $\mathrm{vech}(\uu\uu^\top)$ reduces the feature dimension to $P' = 1+N'+N'(N'+1)/2$, and the calibration requires $K \ge P'$ pilot signals.
Because the dimension $P'$ grows quadratically with $N'$,
the direct use of the Kronecker feature vector may be computationally
prohibitive for a large $N'$.

In such cases, more compact feature representations are preferable.
A random feature map~\cite{Rahimi2007RFF},
$\bm{\varphi}_{\mathrm{RF}}(\uu)
 = \bigl[\cos(\bm{\omega}_j^\top\uu + b_j)\bigr]_{j=1}^{P}$
with i.i.d.\ random frequencies $\bm{\omega}_j$ and phases $b_j$,
approximates a shift-invariant kernel expansion in a $P$-dimensional
space without forming explicit Kronecker products.
Alternatively, the feature map can be parameterized using a neural network,
enabling the end-to-end learning of the identification procedure.
In both cases, Eq.~(\ref{eq:nl_sysid}) remains applicable with
$\bm{\varphi}$ replaced by the chosen feature map, and the reconstruction
requires no explicit physical model of the reservoir.

\section{Linearization of the photonic reservoir response}
\label{sec:photonic_reservoir}

The optical field dynamics inside a microcavity and measured intensity signals 
are described by 
\begin{align}
\dev{\zz(t)}{t}{} = \Am\zz(t)
  + \bb\, E_{\rm in}(t), \\
\xm(t) = |\CC\zz(t)|^2,
\end{align}
where $\zz = (z_1(t), \cdots, z_{M'}(t))^\top \in \Cset^{M'}$ denotes the internal optical field state vector, 
$\Am \in \Cset^{M'\times M'}$ and $\bb \in \Cset^{M'}$ represent the coupling matrix and coupling vector determined by the cavity shape and boundary condition, respectively, $\CC \in \Cset^{M\times M'}$ is the output matrix, $\xm = (x_1(t), \cdots, x_M(t))^\top \in \Rset^M$ represents the measured intensity signals, and $E_{\rm in}(t) = E_0\cos(\alpha u(t) + \phi)$ is the input field modulated by a Mach--Zehnder modulator (MZM) with modulation depth $\alpha$ and bias $\phi$.
From the above equation, the measured intensity $x_m(t)$ at the $m$-th channel can be expressed using the linear response function $H_m(t) = (\CC\,e^{\Am t}\bb)_m$,  
\begin{align}
x_m(t) = \left|\int_{-\infty}^{t} H_m(t-t') E_{\rm in}(t')\,\mathrm{d}t' \right|^2.
\label{eq:intensity_out}
\end{align}
Since $u=0$ before each pilot window, earlier inputs only modify the pilot-independent coefficients of the model below.
In the discretized representation,  
this reads
\begin{align}
\xx_m = |\HH_m \EE_{\rm in}|^2,
\qquad
\xx   = |\HH \EE_{\rm in}|^2,
\label{eq:nonlinear_discrete}
\end{align}
where $\HH_m =\{H_{mnn'}\}$ $(n=1,\ldots,N;\ n'=0,\ldots,N'-1)$ is given by 
\begin{align}
H_{mnn'} =
\begin{cases}
\left(\CC\, e^{\Am (n\Delta t - n'\Delta t')} \bb \right)_{m}\Delta t', & \text{if } n\Delta t \ge n'\Delta t',\\
0, & \text{otherwise},
\end{cases}
\end{align} 
and $\EE_{\rm in}$ collects the input field values sampled at the time interval $\Delta t'$.

Expanding Eq.~(\ref{eq:nonlinear_discrete}) in component form:
\begin{align}
x_m(n\Delta t) &= x_{mn} \nonumber \\
&= \sum_{n',n''} H_{mnn'}\,H_{mnn''}^{*}\,E_{{\rm in},n'}\,E_{{\rm in},n''}^{*},
\label{eq:xmi_expand}
\end{align}
where the cross term $E_{{\rm in},n'}\,E_{{\rm in},n''}^{*}$ is rewritten as,
\begin{align}
&E_{{\rm in},n'}\,E_{{\rm in},n''}^{*}
 = E_0^2 \cos(\alpha u_{n'} + \phi)\cos(\alpha u_{n''} + \phi)
   \nonumber \\
 &= \frac{E_0^2}{2}\Bigl[
     \cos\!\bigl(\alpha(u_{n'} + u_{n''}) + 2\phi\bigr)
   + \cos\!\bigl(\alpha(u_{n'} - u_{n''})\bigr)
   \Bigr].
\label{eq:product_to_sum}
\end{align}
At the standard quadrature bias $\phi = -\pi/4$, the first term yields
$\cos(\alpha(u_{n'} + u_{n''}) - \pi/2) = \sin(\alpha(u_{n'} + u_{n''}))
\approx \alpha(u_{n'} + u_{n''})$ for small $\alpha$,
whereas the second term yields:
\begin{align}
\cos\!\bigl(\alpha(u_{n'} - u_{n''})\bigr)
  \approx 1 - \frac{\alpha^2}{2}(u_{n'} - u_{n''})^2
  \equiv 1 - \alpha^2\Delta_{n'n''},
\label{eq:Deltajj}
\end{align}
where $\Delta_{n'n''} = \tfrac{1}{2}(u_{n'} - u_{n''})^2$. 
From Eqs.~(\ref{eq:product_to_sum}) and (\ref{eq:Deltajj}), Eq.~(\ref{eq:xmi_expand}) can be rewritten as
\begin{align}
x_{mn}
&\approx H_{mn}^{(0)} + \sum_{n'}\tilde{H}_{mnn'}\,u_{n'} \nonumber \\
&{}- \frac{\alpha^2E_0^2}{2}\sum_{n'\neq n''}H_{mnn'}H_{mnn''}^{*}\Delta_{n'n''}
 + O(\alpha^3),
\label{eq:xmi_combined}
\end{align}
where
\begin{align}
H_{mn}^{(0)} &{}= \frac{E_0^2}{2}\Bigl|\sum_{n'}H_{mnn'}\Bigr|^2, \nonumber \\
\tilde{H}_{mnn'} &= \frac{\alpha E_0^2}{2}
   \Bigl(H_{mnn'}\sum_{n''}H_{mnn''}^{*}+H^{*}_{mnn'}\sum_{n''}H_{mnn''}\Bigr).
\end{align}
The third term in Eq.~(\ref{eq:xmi_combined}) is quadratic in $u$ and of order $\alpha^2$; it is therefore negligible compared with the linear term for a small modulation depth. Moreover, because the impulse responses are short, $H_{mnn'}H_{mnn''}^{*} \approx 0$ for $|n'-n''| \gg 1$, so that only a small number of pairs $(n',n'')$ contribute to this term. Neglecting it yields
\begin{align}
x_{mn}
 \approx H_{mn}^{(0)} + \sum_{n'} \tilde{H}_{mnn'}\,u_{n'},
\label{eq:xmi_linear}
\end{align}
where $H_{mn}^{(0)}$ is a static offset.
In matrix form, the offset is incorporated into the augmented input vector
$\tilde{\uu} = (1,\,u_0,\,u_1,\ldots,u_{N'-1})^\top$:
\begin{align}
\xx_m = \tilde{\HH}_m\,\tilde{\uu},
\label{eq:linear_approx}
\end{align}
where $\tilde{\HH}_m = \bigl[\mathbf{h}_m^{(0)},\,(\tilde{H}_{mnn'})_{n,n'}\bigr] \in \Rset^{N\times(N'+1)}$ is the effective intensity transmission matrix, with $\mathbf{h}_m^{(0)} = (H^{(0)}_{m1},\ldots,H^{(0)}_{mN})^\top$.
This is structurally identical to the linear sensing model of
Sec.~\ref{sec:theory}, with the static offset absorbed into the calibrated
reconstruction operator $\hat{\HH}^\dagger$ by appending a constant entry to the measurement vector in Eq.~(\ref{eq:sysid}).

\section{Robustness to timing imperfections and amplitude noise}
\label{sec:robustness}

In this appendix, we analyze the robustness of the reconstruction to two classes of timing imperfections (Appendices~\ref{sec:clockskew} and \ref{sec:jitter}) and to amplitude noise (Appendix~\ref{sec:ampnoise}).

\subsection{Clock skew}
\label{sec:clockskew}

When the $m$-th channel has a deterministic clock offset $\tau_m$,
the measurement vector and sensing sub-matrix are redefined as
\begin{align}
\xx_m &= \bigl(x_m(\Delta t+\tau_m),\;\ldots,\;
         x_m(N\Delta t+\tau_m)\bigr)^\top, \\
\HH_m[n,n'] &{}=
\begin{cases}
h_m(t_{nn'})\,\Delta t', & \text{if } t_{nn'} \ge 0,\\
0, & \text{otherwise},
\end{cases}
\end{align}
where $t_{nn'} = n\Delta t - n'\Delta t' + \tau_m$.
The overall linear model $\xx = \HH\uu$ retains its form with the
skew-adjusted matrices, and the reconstruction $\hat{\uu} = \HH^\dagger\xx$ remains valid. 
Therefore, deterministic clock skew is automatically absorbed
into the calibrated operator $\hat{\HH}^\dagger$.
Because only time-invariant offsets are absorbed, the channels require a common frequency reference, but no phase alignment among their sampling clocks.

\subsection{Jitter noise}
\label{sec:jitter}

Random timing jitter $\delta t_{mn}$ on the $n$-th sample at the $m$-th channel perturbs the measurement by $\delta x_{mn} \approx \dot{x}_m(n\Delta t)\,\delta t_{mn}$, where $\dot{x}_m$ denotes the time derivative of $x_m$. Stacking across all samples, the perturbation vector is
\begin{align}
\delta\xx_m \approx \dot{\xx}_m \odot \delta\boldsymbol{t}_m,
\end{align}
where $\delta\boldsymbol{t}_m = (\delta t_{m1},\ldots,\delta t_{mN})^\top$
and $\odot$ denotes the Hadamard product.
Assuming i.i.d.\ zero-mean jitter with variance $\sigma_t^2$, that is,
$\mathbb{E}[\delta\boldsymbol{t}_m\,\delta\boldsymbol{t}_{m'}^\top]
  = \sigma_t^2\,\delta_{mm'}\,\II_N$,
and applying Bernstein's inequality $|\dot{x}_m(t)| \le 2\pi f_{\rm max}\|x_m\|_\infty \le c\,f_{\rm max}$ with $c \equiv 2\pi\max_m\|x_m\|_\infty$
for $f_{\rm max}$-bandlimited signals,
the covariance of the stacked noise
$\delta\xx = (\delta\xx_1^\top,\ldots,\delta\xx_M^\top)^\top \in \Rset^{MN}$ satisfies
\begin{align}
\mathbb{E}[\delta\xx\,\delta\xx^\top]
\preceq c^2 f_{\rm max}^2\,\sigma_t^2\,\II_{MN}.
\label{eq:noise_cov}
\end{align}

The reconstruction error is $\delta\uu = \HH^\dagger\delta\xx \in \Rset^{N'}$, with
expected squared norm~\cite{Kay1993}
\begin{align}
\mathbb{E}[\|\delta\uu\|^2]
&= \mathrm{Tr}\!\left(
   \HH^\dagger\,\mathbb{E}[\delta\xx\,\delta\xx^\top]\,(\HH^\dagger)^\top
   \right) \nonumber \\
&\le c^2 f_{\rm max}^2\,\sigma_t^2\;
   \mathrm{Tr}\!\left((\HH^\top\HH)^{-1}\right).
\label{eq:mse_bound}
\end{align}
Under the i.i.d.\ channel assumption $\mathbb{E}[\HH_m^\top\HH_m] \equiv \Gamma$, the sum $\HH^\top\HH = \sum_{m=1}^M \HH_m^\top\HH_m$ approaches $M\Gamma$ for $M \gg N_R$.
Consequently, the upper bound of the mean squared error scales as $1/M$:
\begin{align}
\mathbb{E}[\|\delta\uu\|^2]
\lesssim \frac{c^2 f_{\rm max}^2\,\sigma_t^2}{M}\,\mathrm{Tr}(\Gamma^{-1}).
\label{eq:mse_scaling}
\end{align}

\subsection{Amplitude and quantization noise}
\label{sec:ampnoise}

We model the measurements as $\xx_m = \HH_m\uu + \mathbf{n}_m$, where $\mathbf{n}_m$ includes detector, amplifier, ADC, and quantization noise. For a $b_m$-bit ADC with full-scale range $V_m$, the quantization noise has variance $\Delta_m^2/12$ with $\Delta_m = V_m/2^{b_m}$. We assume that the noise is zero mean and independent among channels, with variance $\sigma_m^2$ per sample that may differ from channel to channel; the noise covariance is thus $\boldsymbol{\Sigma}_n = \mathrm{diag}(\sigma_1^2\II_N,\ldots,\sigma_M^2\II_N)$.

Because the calibrated operator $\hat{\HH}^\dagger = \UU^p{\XX^p}^\dagger$ is estimated from pilot measurements that contain the actual noise of each channel, it converges for a large number of pilots to the linear minimum-mean-square-error estimator~\cite{Kay1993}
\begin{align}
\mathbf{W} = \mathbf{C}_u\HH^\top\bigl(\HH\mathbf{C}_u\HH^\top + \boldsymbol{\Sigma}_n\bigr)^{-1},
\label{eq:lmmse}
\end{align}
where $\mathbf{C}_u$ is the covariance of the pilot signals. Assuming white pilots, $\mathbf{C}_u = \sigma_u^2\II$, and test signals with the same statistics, the NMSE of the reconstruction is
\begin{align}
\mathrm{NMSE}
= \frac{1}{N'}\,\mathrm{Tr}\Bigl[\bigl(\II + \sigma_u^2\HH^\top\boldsymbol{\Sigma}_n^{-1}\HH\bigr)^{-1}\Bigr]
= \frac{1}{N'}\sum_{i=0}^{N'-1}\frac{1}{1+\lambda_i},
\label{eq:nmse_lmmse}
\end{align}
where $\lambda_i \ge 0$ are the eigenvalues of
\begin{align}
\sigma_u^2\HH^\top\boldsymbol{\Sigma}_n^{-1}\HH = \sum_{m=1}^{M}\mathrm{SNR}_m\,\HH_m^\top\HH_m,
\label{eq:snr_gram}
\end{align}
where $\mathrm{SNR}_m = \sigma_u^2/\sigma_m^2$ is the signal-to-noise ratio (SNR) at the $m$-th channel.
Because each channel adds a positive semidefinite term to Eq.~(\ref{eq:snr_gram}), the NMSE is non-increasing in $M$: a noisier channel, e.g., one with a coarser quantization, receives a smaller weight but does not degrade the reconstruction. Poorly observed components ($\lambda_i \ll 1$) are not recovered, but their noise is not amplified.

In the high-SNR limit ($\lambda_i \gg 1$, which requires $M \ge N_R$) with equal noise $\sigma^2$ in all channels, $\lambda_i = \mathrm{SNR}\,s_i^2$, where $\mathrm{SNR} = \sigma_u^2/\sigma^2$ and $s_i$ are the singular values of $\HH$, so that
\begin{align}
\mathrm{NMSE} \approx \frac{G}{\mathrm{SNR}},
\qquad
G = \frac{1}{N'}\sum_{i=0}^{N'-1}\frac{1}{s_i^2}.
\label{eq:G}
\end{align}
Compared with an ideal Nyquist-rate ADC ($\mathrm{NMSE} \approx 1/\mathrm{SNR}$), $G$ represents the SNR penalty ($G>1$) or gain ($G<1$) of RSA. 
With the power transfer $\eta_m$ to the $m$-th channel ($\|\HH_m\|_F^2 = \eta_m N$), we have $\sum_i s_i^2 = \|\HH\|_F^2 = N\sum_m\eta_m$, and the inequality between the arithmetic and harmonic means of $\{s_i^2\}_{i=0}^{N'-1}$ gives $G \ge N'/\sum_i s_i^2 = N_R/\sum_m\eta_m$. If each channel is amplified such that $\eta_m = 1$, $G \ge N_R/M$, i.e., $G \ge 1$ for $M = N_R$, and $G$ can be reduced by redundant channels ($M > N_R$). For a passive reservoir without amplification ($\sum_m\eta_m \le 1$), $G \ge N_R$. In both cases, $G$ becomes large when $\HH$ is ill-conditioned.


\bibliography{references_v6}

\end{document}


\raggedbottom

\title{Supplemental Material for\\
``Physical Reservoir Signal Acquisition for Sub-Nyquist Waveform Reconstruction''}

\author{Yuito Ito}
\email{kindaiyuito@stu.kanazawa-u.ac.jp}
\affiliation{%
Graduate School of Natural Science and Technology, Kanazawa University,
Kakuma-machi, Kanazawa, Ishikawa, 920-1192, Japan}

\author{Anas Skalli}
\affiliation{%
Universit\'{e} Marie et Louis Pasteur, CNRS UMR~6174, Institut FEMTO-ST,
15B Avenue Montboucons, Besan\c{c}on, 25000, France}

\author{Tetsuya Asai}
\affiliation{%
Faculty of Information Science and Technology, Hokkaido University,
Kita~14, Nishi~9, Kita-ku, Sapporo, Hokkaido, 060-0814, Japan}

\author{Satoshi Sunada}
\email{sunada@se.kanazawa-u.ac.jp}
\affiliation{%
Faculty of Mechanical Engineering, Institute of Science and Engineering,
Kanazawa University, Kakuma-machi, Kanazawa, Ishikawa, 920-1192, Japan}

\date{\today}

\maketitle

\section{Reconstruction error versus the number of channels}
\label{sec:S_nmse_M}

Table~\ref{tab:nmse_M} lists the NMSE for broadband random waveforms (Fig.~6 of the main text) for $M=1$--$6$. Because the oscilloscope has only four channels, the results for $M\geq5$ were obtained pseudo-simultaneously: the same input waveform sequence was applied repeatedly, and the reservoir outputs recorded in separate acquisitions were combined. In the redundant regime ($M>N_R=4$), the NMSE decreases further owing to the redundancy of the measurements, consistent with the noise analysis in Appendix~C\,3 of the main text. However, the decrease saturates at $\approx0.02$ for $M=5$ and 6, which is attributed to residual errors such as calibration errors.

\begin{table}[H]
\caption{NMSE of the reconstruction of broadband random waveforms versus $M$ (mean and standard deviation over 30 test waveforms).}
\label{tab:nmse_M}
\begin{ruledtabular}
\begin{tabular}{ccc}
$M$ & Mean & Standard deviation \\
\hline
1 & 0.569 & 0.050 \\
2 & 0.232 & 0.024 \\
3 & 0.084 & 0.011 \\
4 & 0.0319 & 0.0057 \\
5 & 0.0220 & 0.0038 \\
6 & 0.0205 & 0.0036 \\
\end{tabular}
\end{ruledtabular}
\end{table}

\section{PAM4 reconstruction versus the number of channels}
\label{sec:S_pam4}

Table~\ref{tab:ber_M} lists the bit error rate (BER) and NMSE of the reconstructed PAM4 signal [Fig.~8(c) of the main text] for $M=1$--$6$. As in Sec.~\ref{sec:S_nmse_M}, the results for $M\geq5$ were obtained by applying the same PAM4 sequence repeatedly and combining the reservoir outputs recorded in separate acquisitions.

\begin{table}[H]
\caption{BER and NMSE of the reconstructed 25-GBd PAM4 signal versus $M$, evaluated over all $1.8\times10^{4}$ bits. Uncertainties are the standard errors of the mean over 45 segments of 200 symbols.}
\label{tab:ber_M}
\begin{ruledtabular}
\begin{tabular}{cccc}
$M$ & Bit errors & BER & NMSE \\
\hline
1 & 6839 & $(3.799 \pm 0.042)\times 10^{-1}$ & $0.7588 \pm 0.0054$ \\
2 & 4423 & $(2.457 \pm 0.028)\times 10^{-1}$ & $0.5222 \pm 0.0055$ \\
3 & 2166 & $(1.203 \pm 0.019)\times 10^{-1}$ & $0.2826 \pm 0.0070$ \\
4 & 68 & $(3.78 \pm 0.86)\times 10^{-3}$ & $0.0201 \pm 0.0014$ \\
5 & 46 & $(2.56 \pm 0.39)\times 10^{-3}$ & $0.0187 \pm 0.0013$ \\
6 & 44 & $(2.44 \pm 0.39)\times 10^{-3}$ & $0.0186 \pm 0.0013$ \\
\end{tabular}
\end{ruledtabular}
\end{table}

\section{Different reservoir realizations}
\label{sec:S_wavelength}

\begin{figure}[H]
\centering
\includegraphics[width=0.55\linewidth]{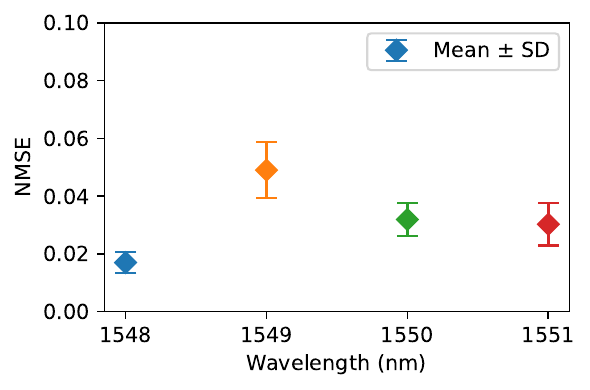}
\caption{NMSE of the reconstruction for different operating wavelengths, i.e., different reservoir realizations. Symbols and error bars denote the mean and standard deviation over 30 test waveforms ($M=4$).}
\label{fig:wavelength}
\end{figure}

Operating the chip at different wavelengths provides different reservoir realizations. Figure~\ref{fig:wavelength} shows the NMSE for four wavelengths between 1548 and 1551~nm, each with a separate calibration ($M=4$). The NMSE ranges from 0.017 to 0.049, indicating that accurate reconstruction does not rely on a particular realization.

